\documentclass[11pt,a4paper]{article}

\usepackage[a4paper,margin=1in,includehead,headheight=60pt,headsep=12pt]{geometry}
\usepackage{multirow}
\newcommand*{\doi}[1]{%
   \href{https://doi.org/#1}{doi:~{\small\nolinkurl{#1}}}}
\usepackage{graphicx}
\usepackage{fancyhdr}
\usepackage{tikz}
\usepackage[hidelinks]{hyperref}
\usepackage[numbers]{natbib}
\usepackage{pgffor} 
\usepackage{minted} 
\usepackage{forest}
\usepackage{booktabs}

\usepackage{longtable}
\fancypagestyle{firstpage}{
  \fancyhf{}

}

\usepackage{subcaption}  
\usepackage{caption}     

\usepackage{tikz}
\usepackage{pgfplots}
\pgfplotsset{compat=1.18}
\usepackage{xspace}
\usepackage{dirtree}
\definecolor{folderbg}{RGB}{124,166,198}
\definecolor{folderborder}{RGB}{110,144,169}
\def\Size{4pt}
\tikzset{
  folder/.pic={
    \filldraw[draw=folderborder,top color=folderbg!50,bottom color=folderbg]
      (-1.05*\Size,0.2\Size+5pt) rectangle ++(.75*\Size,-0.2\Size-5pt);  
    \filldraw[draw=folderborder,top color=folderbg!50,bottom color=folderbg]
      (-1.15*\Size,-\Size) rectangle (1.15*\Size,\Size);
  }
}

\author{Anna Zaborowska (CERN) and Peter McKeown (CERN)}
\date{\today}
\newcommand{\dataset}{The step2point\xspace}
\title{\dataset dataset}
\usepackage{graphicx}

\begin{document}
\maketitle

\begin{abstract}
This dataset contains a detailed simulation output that allows the construction and study of different data representations for electromagnetic and hadronic showers in calorimeters. It is published so that optimal data representations can be studied, with the ultimate goal of constructing a general tool that takes detailed simulation output and translates it into an optimal representation that can serve as the input to surrogate simulators based on generative models.
\end{abstract}

\section{Background \& Summary}

The simulation of particle transport in detectors can be very CPU intensive, and hence many studies propose(d) different machine learning (ML) surrogates to replace the most computationally expensive part of a full detector simulation: the simulation of showers in calorimeters. The recent review published as a result of the CaloChallenge~\cite{Krause:2024avx} describes the landscape of models submitted to the challenge.

A typical simulation takes all deposits from the various steps that fall within a single cell and processes them to create a response. Any fast shower simulation model could therefore attempt to directly parameterise the detector response. However, this poses several challenges, from the strong dependency on the incident angle to the detector surface, to the (often) low granularity images of showers and the issues this creates for placement back into the detector geometry. For this reason, most studies take more granular data from the simulation, with many exploiting the natural symmetry of showers (considering a distinct direction for the axis of propagation of the incident particle).

Most commonly simulation data is (cylindrically) voxelised, either in a regular grid structure, or with voxels created based on average energy densities. Many recent approaches turned towards point cloud representations, with either indirect translation from the voxelised representation (which is sub-optimal), or with some detector-specific method applied to group energy deposits in the vicinity (e.g. division of detector cells into smaller ones)~\cite{Buhmann:2023bwk, Buss:2025cyw}. This step is necessary given the prohibitively large number of simulation steps present in showers at the energies of interest. There is a clear motivation for a study into optimal data representations and for a generic tool that can cluster the detailed simulation into an optimised data format to be input to this class of generative models. The task for a given detector is therefore to find the minimum set of points, created by clustering individual simulation steps, that does not disturb the relevant physics observables at the level of the detector readout.

While there is much research around the application of machine learning models to the parameterisation of electromagnetic (EM) showers, there are few activities focusing on the parameterisation of hadronic showers and that is where we believe the current research should focus.

\section{Methods}

This dataset contains the detector response to a single incident particle. It was produced using the OpenDataDetector~\cite{odd}, with simulation performed with DD4hep~\cite{Gaede:2020tui} via the standard key4hep stack~\cite{Carceller:2025ydg}.

The output of the simulation is stored in the EDM4hep~\cite{Gaede:2022leb} format in ROOT files that contain more information than is needed, so the translation to a dedicated \texttt{HDF5} file is done with a lightweight set of scripts within the \texttt{step2point} repository~\cite{step2point}. 

\section*{Nomenclature}

The following description is a simplification to illustrate the main concepts and terminology that would enable an understanding of the dataset.

When a particle enters the detector, it travels through the detector and interacts with its materials, depositing energy and creating secondary particles. The energy that the detector measures and its placement (in space or time) can be called a \textbf{detector response}. This is measured in a certain structure, with the detector divided into sub-detectors, layers, cells, etc. which will be referred to as the \textbf{detector readout}. The smallest physical unit of readout is a \textbf{cell}. An \textbf{event} is a detector response to a certain input (\textbf{primary particles}), which is typically complex, involves many particles and corresponds to the modelling of a particle collision, but in this instance the focus is solely on single particle inputs. The particles that enter the sub-detector called the \textbf{calorimeter}, whose purpose is to measure the energy of incident particles by absorbing (stopping) them, create \textbf{cascades} of secondary particles called \textbf{showers}. If the incident particle is an electron e$^-$, positron e$^+$, or photon $\gamma$, almost exclusively electromagnetic interactions with matter occur, hence the showers they produce are called \textbf{electromagnetic (EM) showers}. If a hadron interacts with the calorimeter (e.g. a pion $\pi^\pm$ or proton p), it can interact via the strong interaction, as well as by electromagnetic interactions, forming a \textbf{hadronic shower}. Hadronic showers are significantly more complex and exhibit larger variations than the EM showers. In the \textbf{full simulation} performed with Geant4~\cite{Geant4}, each simulated particle (\textbf{MC particle} = Monte Carlo particle) traverses the detector in little \textbf{steps}, and at each one there could have been energy deposited and/or secondary particles created.

\section*{Dataset: file structure}

\begin{figure}[!ht]
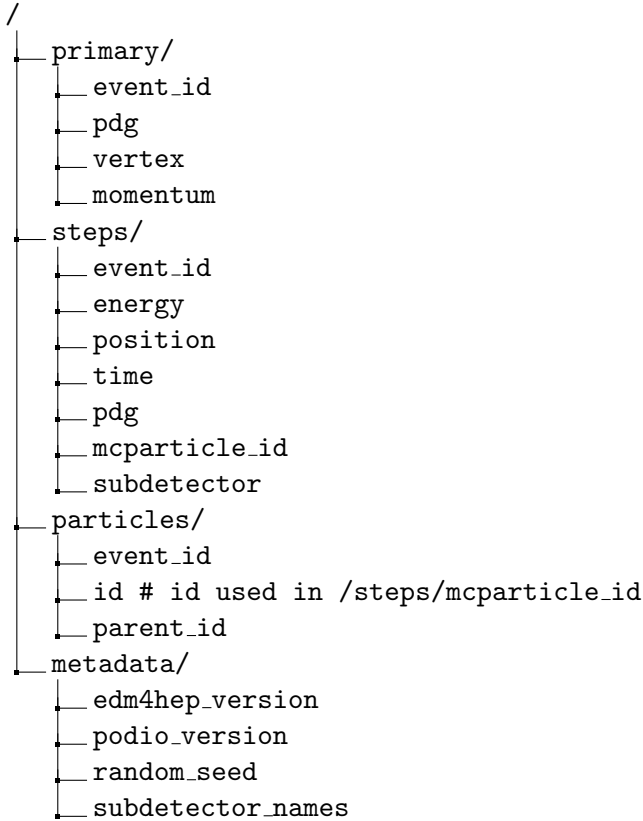

\dirtree{%
.1 /.
.2 primary/.
.3 event\_id.
.3 pdg.
.3 vertex.
.3 momentum.
.2 steps/.
.3 event\_id.
.3 energy.
.3 position.
.3 time.
.3 pdg.
.3 mcparticle\_id.
.3 subdetector.
.2 particles/.
.3 event\_id.
.3 id \# id used in /steps/mcparticle\_id.
.3 parent\_id.
.2 metadata/.
.3 edm4hep\_version.
.3 podio\_version.
.3 random\_seed.  
.3 subdetector\_names.
}
\caption{The structure of the HDF5 files. Each of the 3 groups (\texttt{primary}, \texttt{steps}, \texttt{particles}) contains datasets, while \texttt{metadata} contains information necessary to reproduce the datasets.}
\label{fig:structure}
\end{figure}

The file structure of this dataset is depicted in Fig.~\ref{fig:structure}. There are three groups, each collecting different types of information. Details of each group are presented after the following short description and explanation of how they can be linked with each other. The \textbf{\texttt{primary}} group represents the information about the primary particles, with one entry per event in this dataset. The \textbf{\texttt{steps}} group represents the energy deposits that were made during the simulation, and they can be connected to the \texttt{primary} information via the event identifier \texttt{event\_id}. Such a flat structure was chosen as the number of deposits per event varies, and the overhead of the repeated entries e.g. in the event identifier is negligible thanks to the compression of hdf5 files. In addition to those two groups, a third one called \textbf{\texttt{particles}} is stored to allow the lineage of particles to be recreated, and to link the energy deposits to the particles that created them. First, a matching of an event via the event identifier \texttt{event\_id} would need to be performend, and afterwards the \texttt{mcparticle\_id} from \texttt{steps} could be matched with the \texttt{id} from \texttt{particles}.

\section*{\texttt{primary} group}

This group stores one entry per primary particle per event. This dataset contains only single-particle events, so the number of entries in each dataset should be equal to the number of events,~$N$.

\begin{tabular}{@{}llll@{}}
\toprule
Dataset & Type & Shape & Description \\
\midrule
event\_id  & int32      & (N) & Event index \\
pdg       & int32      & (N) & PDG\footnote{PDG is a unique identifier of a particle type that can be encoded using~\cite{}} code of the primary particle \\
vertex    & float32[3] & (N) & global x, y, z position of particle vertex in mm \\
momentum  & float32[3] & (N) & px, py, pz momentum of particle vertex in GeV/$c$ \\
\bottomrule
\end{tabular}

\section*{\texttt{steps} group}

Each entry represents a simulation step with energy deposition. Assuming that $N$ is the number of events, $M_i$ is the number of steps in $i$-th event, and $M$ is the number of steps across all events:

\[
M = \sum_{i=1}^{N} M_i
\]

\begin{tabular}{@{}llll@{}}
\toprule
Dataset & Type & Shape & Description \\
\midrule
event\_id      & int32      & (M,)   & Event number the step belongs to \\
energy         & float32    & (M,)   & Energy deposited in step (MeV) \\
position       & float32[3] & (M,3)  & global x, y, z position in mm \\
time           & float32    & (M,)   & Timestamp of step (ns) \\
pdg            & int32      & (M,)   & PDG code of contributing particle \\
mcparticle\_id & int32      & (M,)   & ID/index of contributing MCParticle, \\
cell\_id       & uint64     & (M,)   & ID/index of the cell of detector to which step belongs \\
subdetector    & uint8      & (M,)   & Index into /metadata/subdetector\_names \\
\bottomrule
\end{tabular}

\medskip

\noindent\texttt{cell\_id} is a long integer that stores bitfield information about the hierarchy of volumes in which energy was deposited. It is linked to the detector which it describes. For the Open Data Detector it is defined as:

\texttt{"system:5,side:2,module:8,stave:4,layer:9,submodule:4,x:32:-16,y:-16"},

\noindent which reads: Out of my 64 bits in the ID, dedicate the 5 lowest to encode the ID of a system, the next 2 bits for the side, etc. The description \texttt{x:32:-16} means that the encoding field \texttt{x} starts at the 32-nd bit, is given a length of 16 bits and uses signed integers for this ID (by default the field ID is non-negative). To help encoding the bitfield, \texttt{utils/bitfield.py} from \texttt{step2point} can be used.

\section*{\texttt{particles} group}

This group stores at least one entry per Monte Carlo particle per event (more than one if there are multiple parents). All particles from the EDM4hep format will be stored, including the primaries as well as the secondaries created in the simulation which are retained in the output (e.g. because they created an energy deposit of interest in the detector). This means that \textbf{particles that did not create a deposit are not stored}. Each row corresponds to a (child particle, parent particle) pair within a given event. If a particle has multiple parents, it appears multiple times with different parent\_ids. Particles with no parent have parent\_id = -1 (primary particles).

Assuming that $N$ is the number of events, $P_i$ is the number of particles simulated and kept in $i$-th event, and $P$ is the number of particles across all events:

\[
P = \sum_{i=1}^{N} P_i
\]

\begin{tabular}{@{}llll@{}}
\toprule
Dataset & Type & Shape & Description \\
\midrule
event\_id   & int32 & (P) & Event index \\
id          & int32 & (P) & Particle ID (matches mcparticle\_id in /steps) \\
parent\_id  & int32 & (P) & Parent particle ID; -1 if no parent (primary)\\
\bottomrule
\end{tabular}

\section*{Metadata}

Information about the dataset necessary to reproduce the files. Please note that the random seed was only stored for the discrete part of this dataset. Table~\ref{tab:seeds} shows the values of the random seeds necessary to reproduce the continuous part of this dataset (if simulation is to be rerun).

\begin{tabular}{@{}lll@{}}
\toprule
Path & Type & Description \\
\midrule
subdetector\_names & str[]  & List of subdetector names as strings \\
edm4hep\_version             & str    & Version of podio stored as attribute \\
podio\_version               & str    & Version of EDM4hep stored as attribute \\
random\_seed                 & uint64 & Random seed passed to simulation   \\
\bottomrule
\end{tabular}



\section{Data Records}

The dataset is published on Zenodo~\cite{step2point_zenodo}. There are two types of simulated samples: those with discrete and those with continuous properties of the primary particles. All of them are created for 3 particle types: photons (called gammas), protons (positive charge), and pions with negative charge.

\begin{figure}[htbp]
        \centering
    \includegraphics[width=0.4\linewidth]{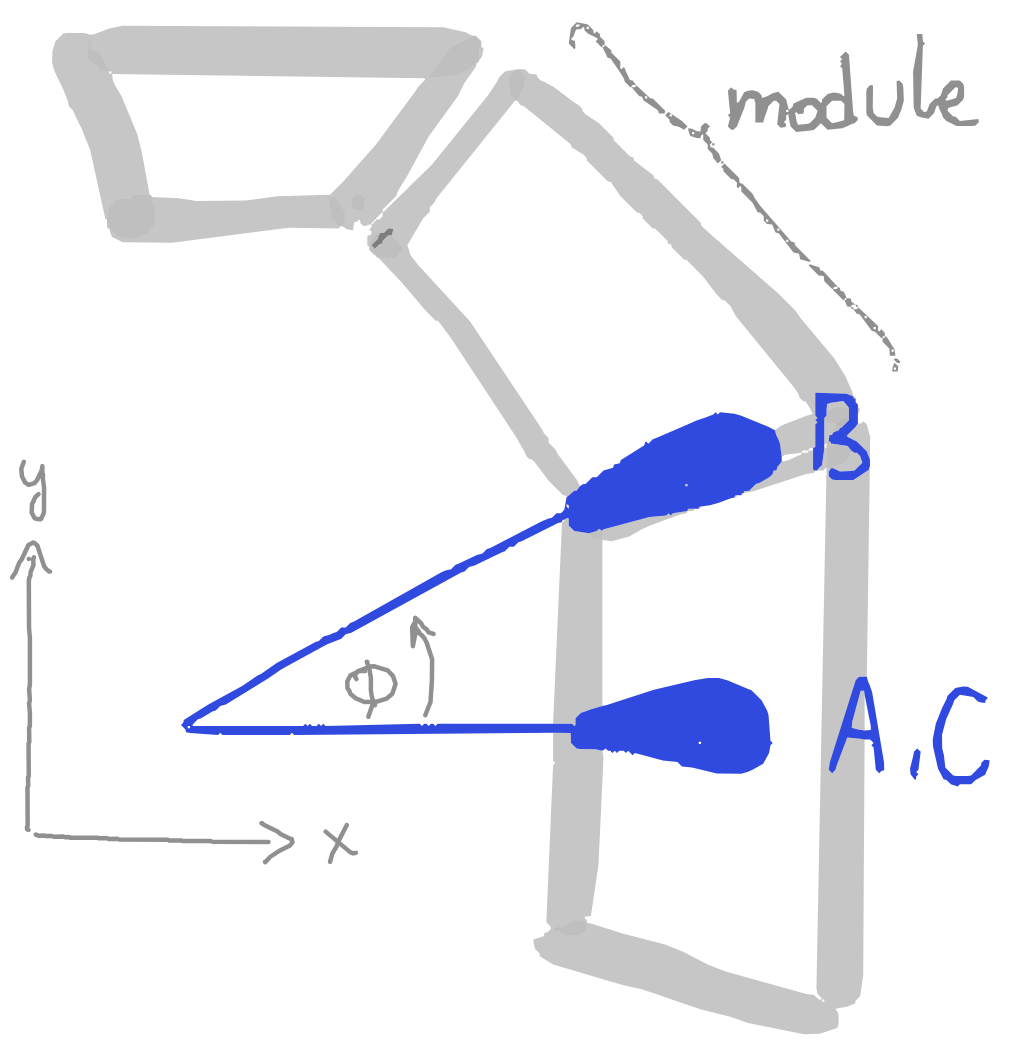}~
    \includegraphics[width=0.6\linewidth]{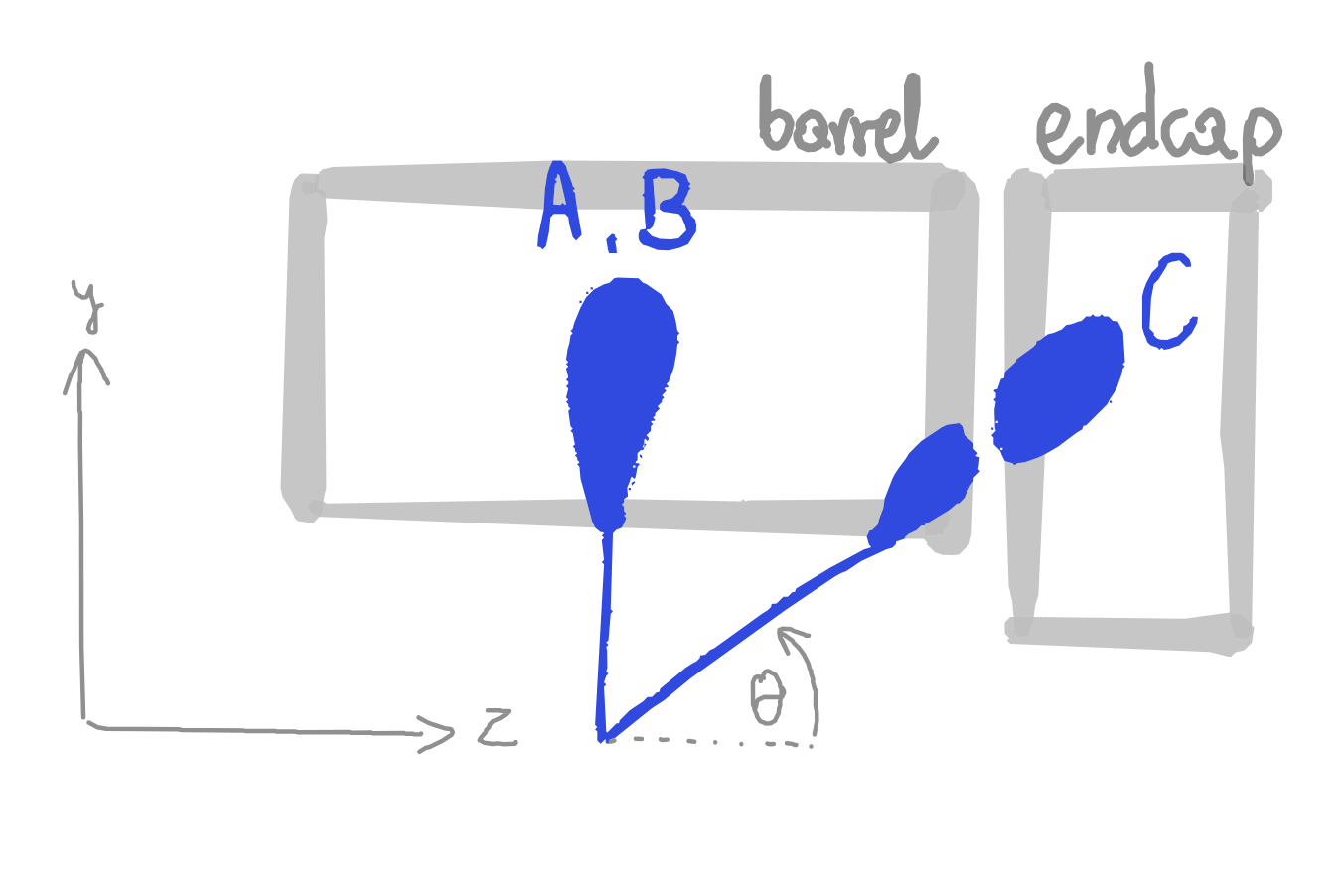}
        \caption{A sketch of the detector with the directions and kinematic scenarios explained. The azimuthal $\phi$ and polar $\theta$ angles are depicted, together with a definition of the regions of the detector: a module is illustrated (in the XY cross-section), as are the barrel and endcap regions of the detector. All positions are expressed in global coordinates, and angles are relative to the centre of the detector.}
        \label{fig:direction}
\end{figure}

\begin{figure}[htbp]
        \centering
        \includegraphics[width=0.8\textwidth]{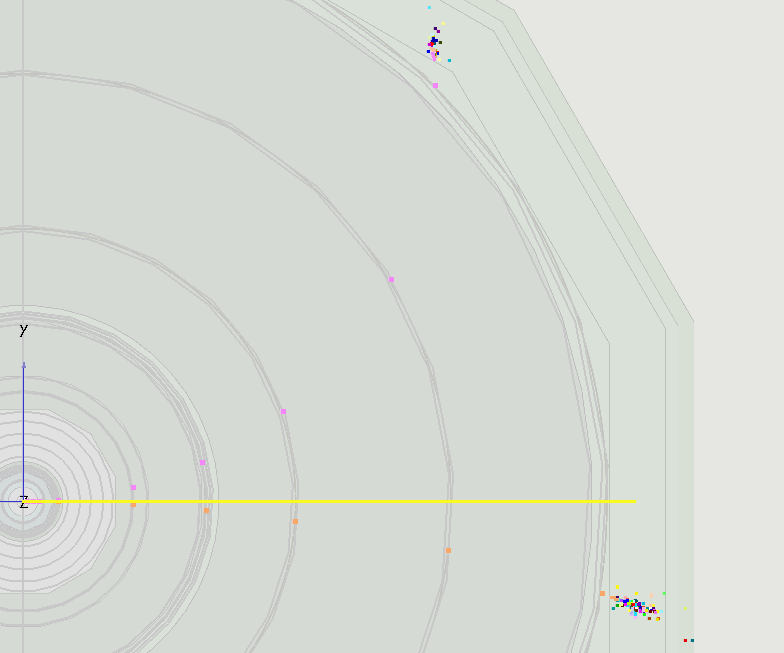}
        \caption{An event display for a 5\,GeV photon that interacted early in the tracking detector. The photon converts into an electron-positron pair, with the passage of those leaving signals in the tracking layers, depicted by the pink and orange markers. The resulting positive and negative particles are bent in opposite directions in the magnetic field of the detector and once they enter the calorimeter they create showers (multi-coloured markers). The direction of the incident photon is depicted in yellow.}
        \label{fig:paircreation}
\end{figure}

The continuous dataset is split into 5 files, with each file containing 10`000 showers. The primary particle energies range from 0.1\,GeV to 100\,GeV and are distributed uniformly. The direction of the particles is uniformly sampled from $\theta=6^\circ$ to $\theta=174^\circ$, and the  azimuthal angle is sampled uniformly across the full azimuthal range ($-\pi$ to $\pi$). This represents particles travelling in almost all directions, with the exception of the beampipe direction. The production vertex of each particle is at the centre of the detector, $(x,y,z)=(0,0,0)$. A summary of the incident particle properties can be found in Tab.~\ref{tab:dataset:cont} and the ranges are depicted in Fig.~\ref{fig:direction}. This means that particles traverse the beampipe and the tracking sub-detector before they enter the calorimeter system. There may be some initial interactions, including energy depositions and scattering, with the example of a photon conversion (a photon converting into an electron-positron pair) shown in Fig.~\ref{fig:paircreation}, completely changing the expectation of a single cluster in the detector. In such cases the particle that enters the calorimeter (in this case an electron/positron) should be considered as the particle that initiates the shower.

\begin{table}[h!]
\centering
\begin{tabular}{cc|c|c}
\hline
\multicolumn{2}{c|}{\textbf{Incident particle property}} & \textbf{Unit} & \textbf{Values}\\
\hline
\multicolumn{2}{c|}{particle type (and PDG)} &  & gamma(22), proton(2212), pion(-211) \\
\multicolumn{2}{c|}{Energy} & GeV & 0.1--100 \\
\multicolumn{2}{c|}{Azimuthal angle ($\phi$)} & deg & 0 -- 360 \\
\multicolumn{2}{c|}{Polar angle ($\theta$)} & deg & 6 -- 174\\
\hline
\multirow{3}*{Vertex position} 
 & $x$ & mm & 0 \\
 & $y$ & mm & 0  \\
 & $z$ & mm & 0  \\
\hline
\end{tabular}
\caption{Incident particle properties and vertex positions for the simulation of the continuous part of the dataset. For each of the three incident particles (photon, proton, pion) particles are simulated with a continuous range of energies and angles ($\phi, \theta$).}\label{tab:dataset:cont}
\end{table}

The discrete dataset has been produced to create a controlled environment, with incident particles produced right in front of the calorimeter system, which mitigates the visible effects of particle interaction with the sub-detector(s) in front of the calorimetrs. The discrete energies range from low to high energies (0.1, 1, 10, 100 GeV), and the direction of the particles in the detector represents 3 different regions, as presented in Fig.~\ref{fig:direction}: in the middle of the detector and perpendicular to the calorimeter face, in between the modules of the detector (corner of a polygon in the transverse cross-section), and in between the so-called barrel and endcap, representing a gap in coverage between the parts of calorimeters. For the latter, a different choice is made for photons and hadrons, as photons interact with the electromagnetic calorimeter, while for hadrons the gap in the hadronic calorimeter system is of relevance. Table X summarises the choice of energies, angles, and production vertices (chosen based on the angles and inner radius of the electromagnetic calorimeter, as they should point to the origin).

\begin{table}[h!]
\centering
\begin{tabular}{cc|c|c|c|c}
\hline
\multicolumn{2}{c|}{\textbf{Incident particle property}} & \textbf{Unit} & \multicolumn{3}{c}{\textbf{Values}}\\
\hline
\multicolumn{2}{c|}{particle type (and PDG)} &  & \multicolumn{3}{c}{gamma(22), proton(2212), pion(-211)} \\
\multicolumn{2}{c|}{Energy} & GeV & \multicolumn{3}{c}{0.1, 1, 10, 100} \\
\hline
\multicolumn{3}{c|}{\textbf{Kinematic scenario, Fig.~\ref{fig:direction}}}  & \textbf{A} & \textbf{B} & \textbf{C} \\
\hline
\multicolumn{2}{c|}{Azimuthal angle ($\phi$)} & deg & 0 & 11.25 & 0 \\
\multicolumn{2}{c|}{Polar angle ($\theta$)} & deg & 90 & 90 & 23 ($\gamma$) or 25 (hadrons) \\
\hline
\multirow{3}*{Vertex position} 
 & $x$ & mm & 1250 & 1250 & 1250 \\
 & $y$ & mm & 0 & 248 & 0 \\
 & $z$ & mm & 0 & 0 & 2681 \\
\hline
\end{tabular}
\caption{Incident particle properties and vertex positions for the simulation of the discrete part of the dataset. For each of the three incident particles, 4 energy values are simulated at three different kinematic scenarios, depicted in Fig.~\ref{fig:direction}.}\label{tab:dataset:discrete}
\end{table}

\begin{figure}[htbp]
        \centering
        \includegraphics[width=1\textwidth,trim={0 20pt 0 35pt},clip]{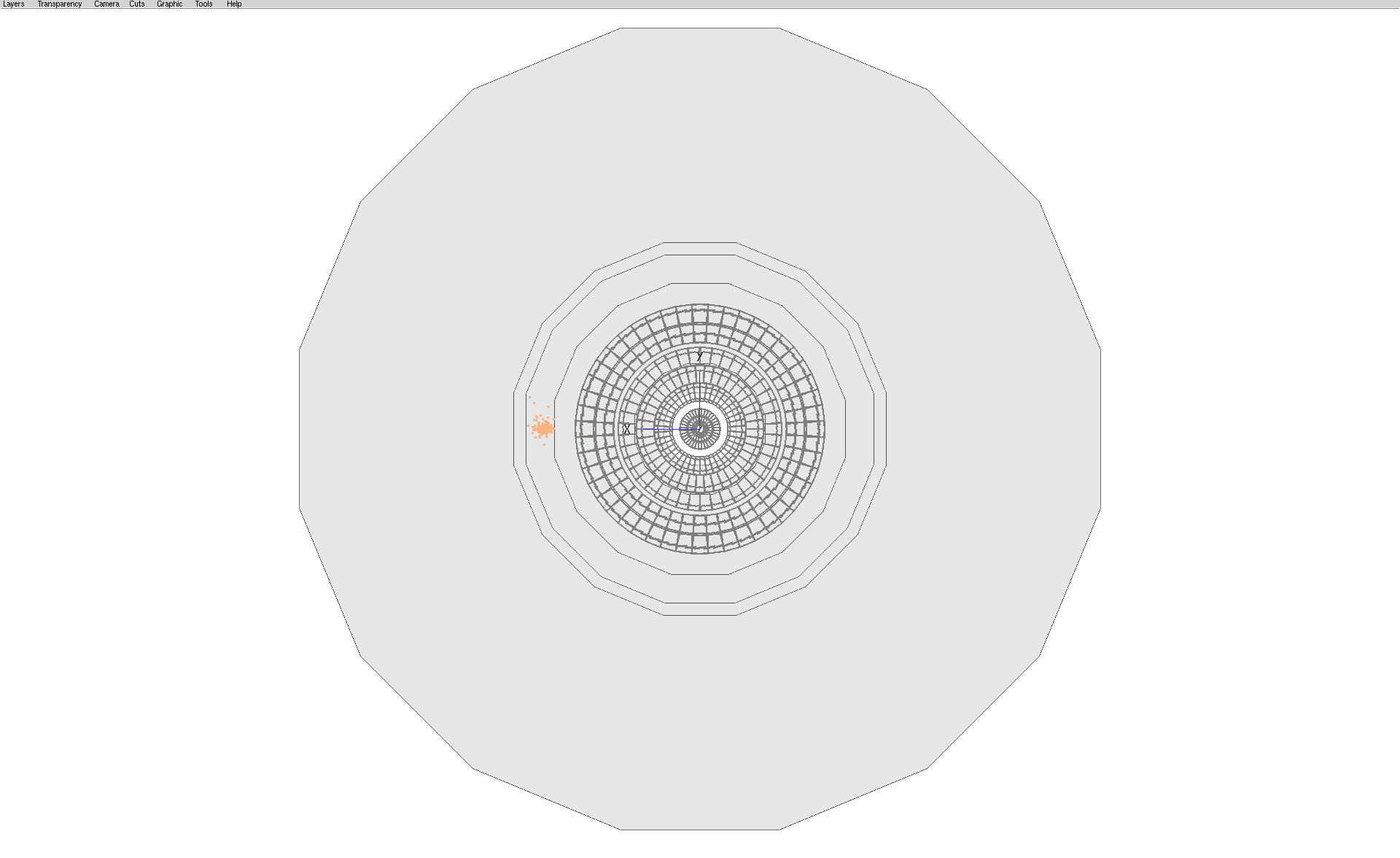}
        \caption{An event visualisation of a 10 GeV photon that enters the detector at $\theta=90^\circ$ and  $\phi=0^\circ$, which corresponds to the middle of the detector module, and is oriented perpendicularly to the calorimeter layers.}
        \label{fig:gamma:phi0}
\end{figure}
\begin{figure}[htbp]
        \centering
        \includegraphics[width=1\textwidth,trim={0 20pt 0 35pt},clip]{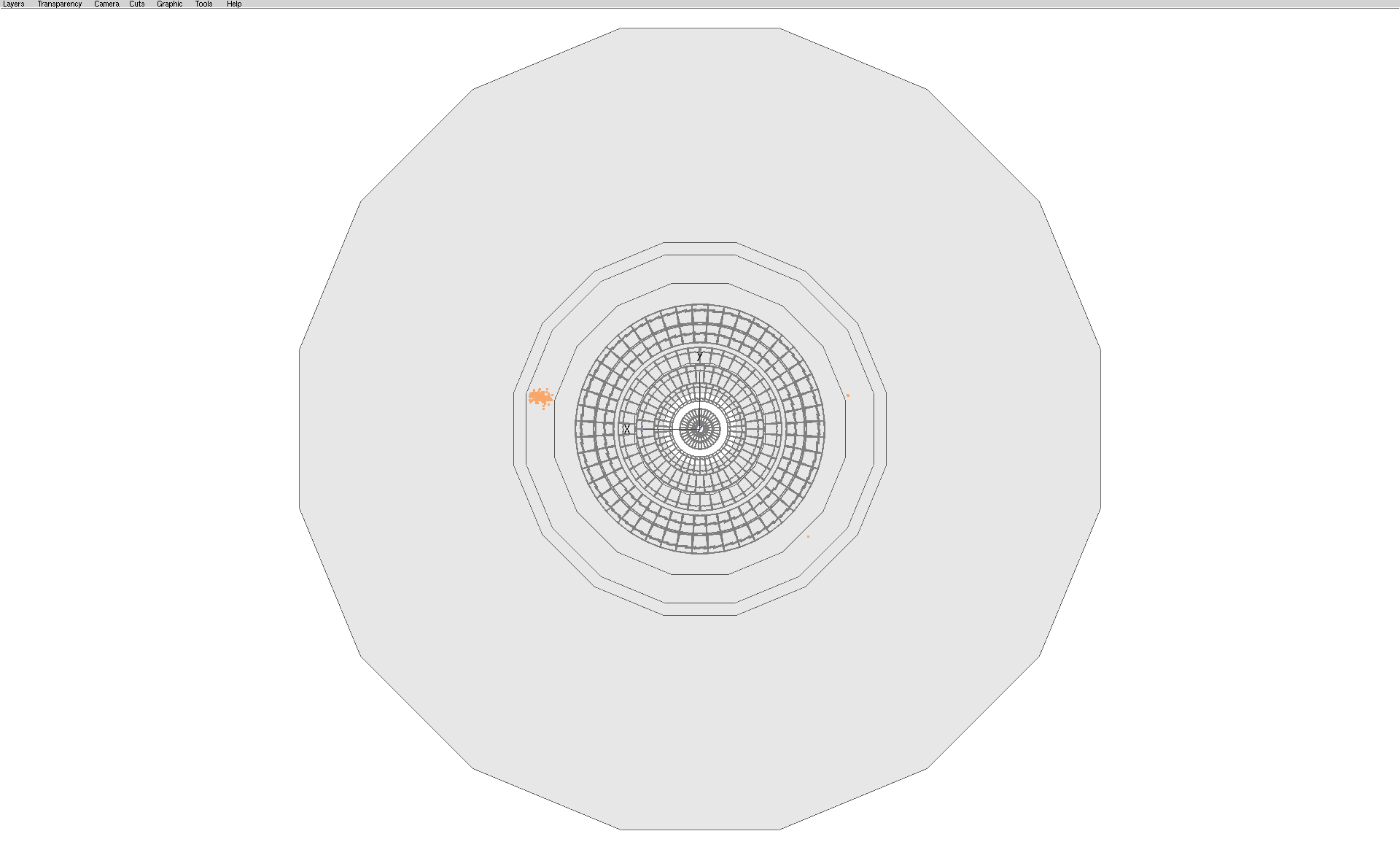}
        \caption{An event visualisation of a 10 GeV photon that enters the detector at $\theta=90^\circ$ and  $\phi=11.25^\circ$, which corresponds to the corner between the detector modules.}
        \label{fig:gamma:phi11}
\end{figure}
\begin{figure}[htbp]
        \centering
        \includegraphics[width=1.\textwidth,trim={400pt 20pt 400pt 50pt},clip]{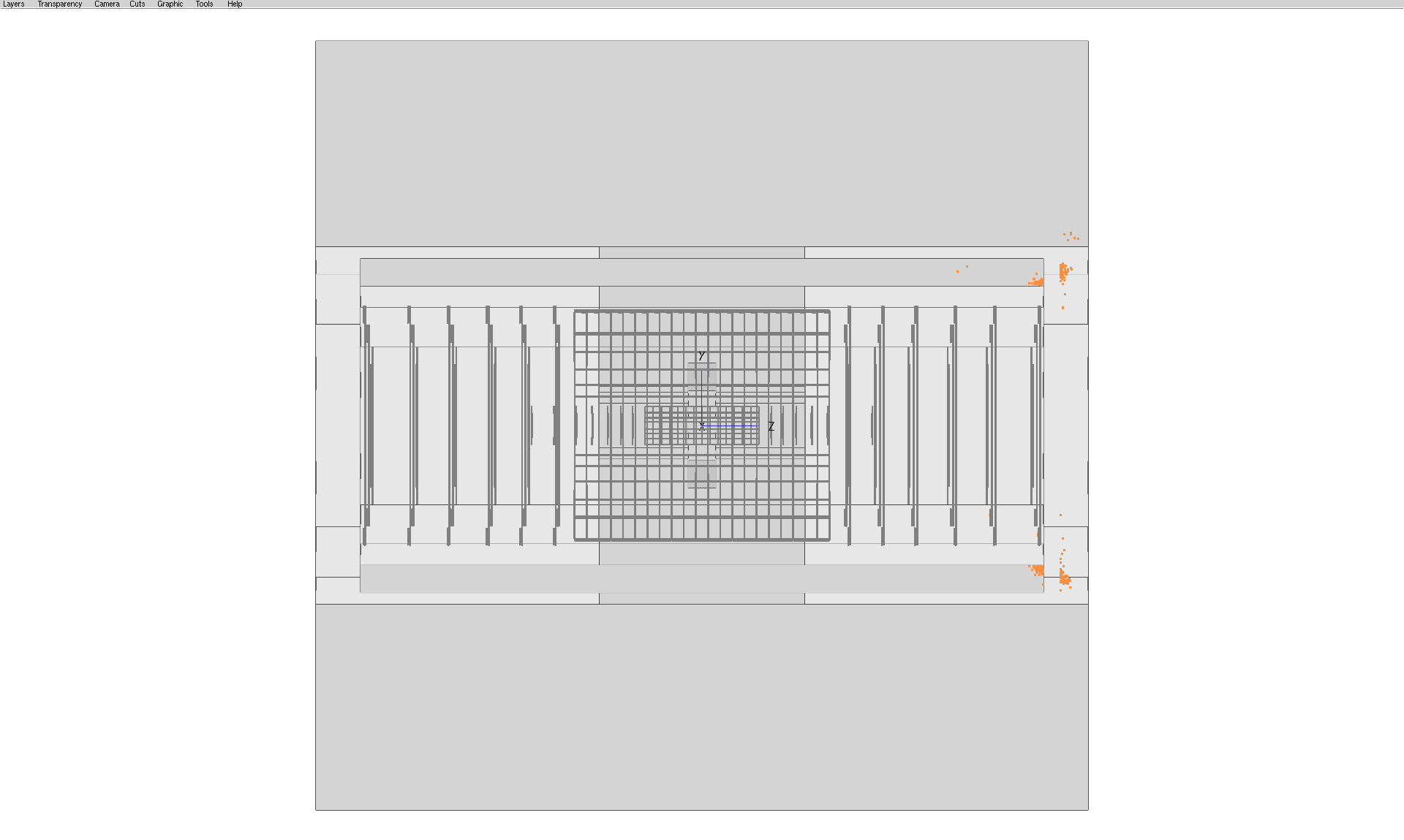}
        \caption{An event visualisation of a 10 GeV photon that enters the detector at $\theta=23^\circ$ and  $\phi=0^\circ$, which corresponds to the transition between the barrel and endcap (the endcap is not displayed). The showers seems to leave deposition in two regions of the detector (bottom and top), but that is an artifact of how $\phi$ is displayed, as part of shower has low values, while the other part has values close to $360^\circ$.}
        \label{fig:gamma:theta23}
\end{figure}
\begin{figure}[htbp]
        \centering
        \includegraphics[width=1\textwidth,trim={0 20pt 0 35pt},clip]{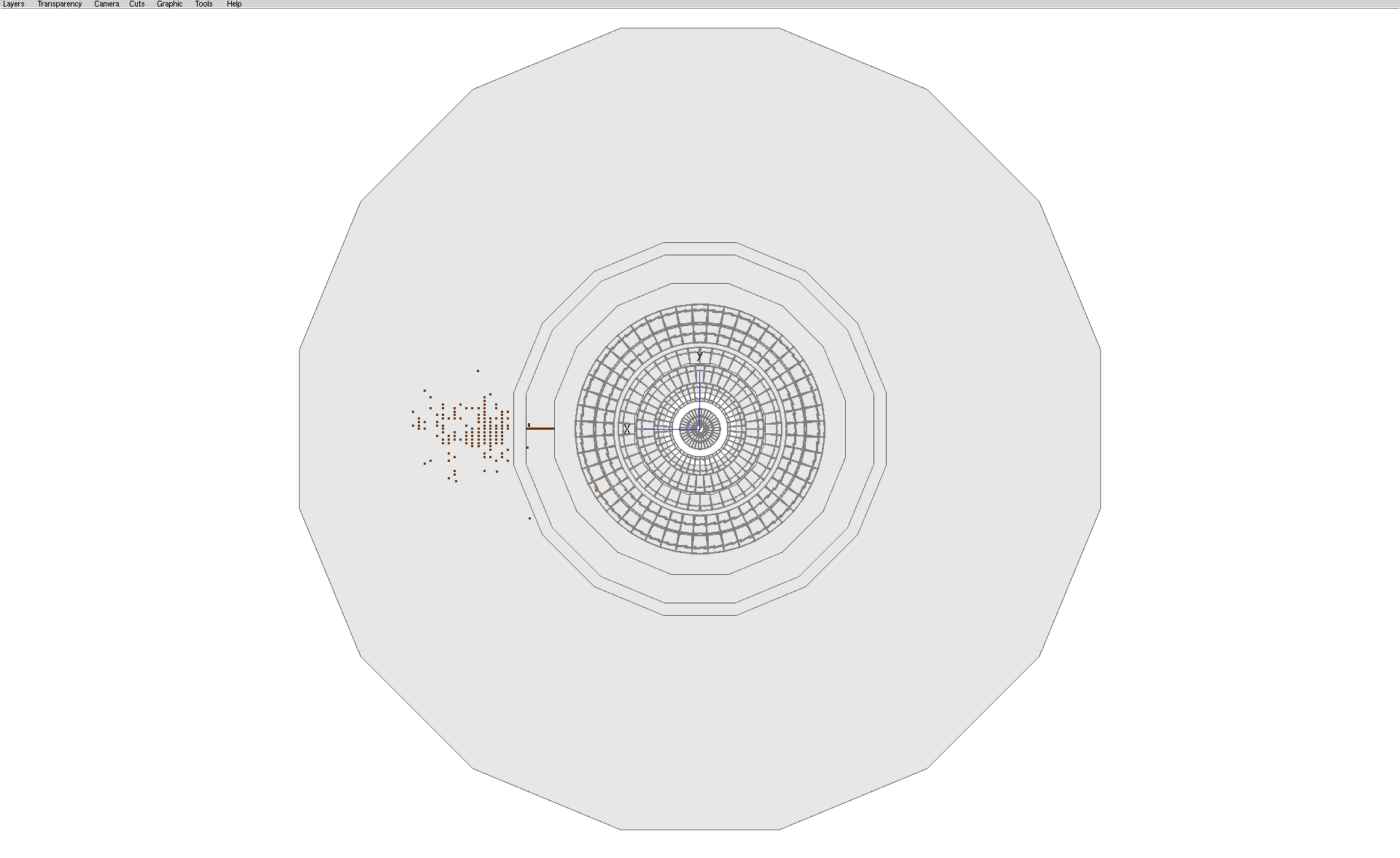}
        \caption{An event visualisation of a 10 GeV proton that enters the detector at $\theta=90^\circ$ and  $\phi=0^\circ$. The proton traverses almost the entire region of the electromagnetic calorimeter (leaving a track), before initiating a shower.}
        \label{fig:proton:phi0}
\end{figure}
\begin{figure}[htbp]
        \centering
        \includegraphics[width=1\textwidth,trim={0 20pt 0 35pt},clip]{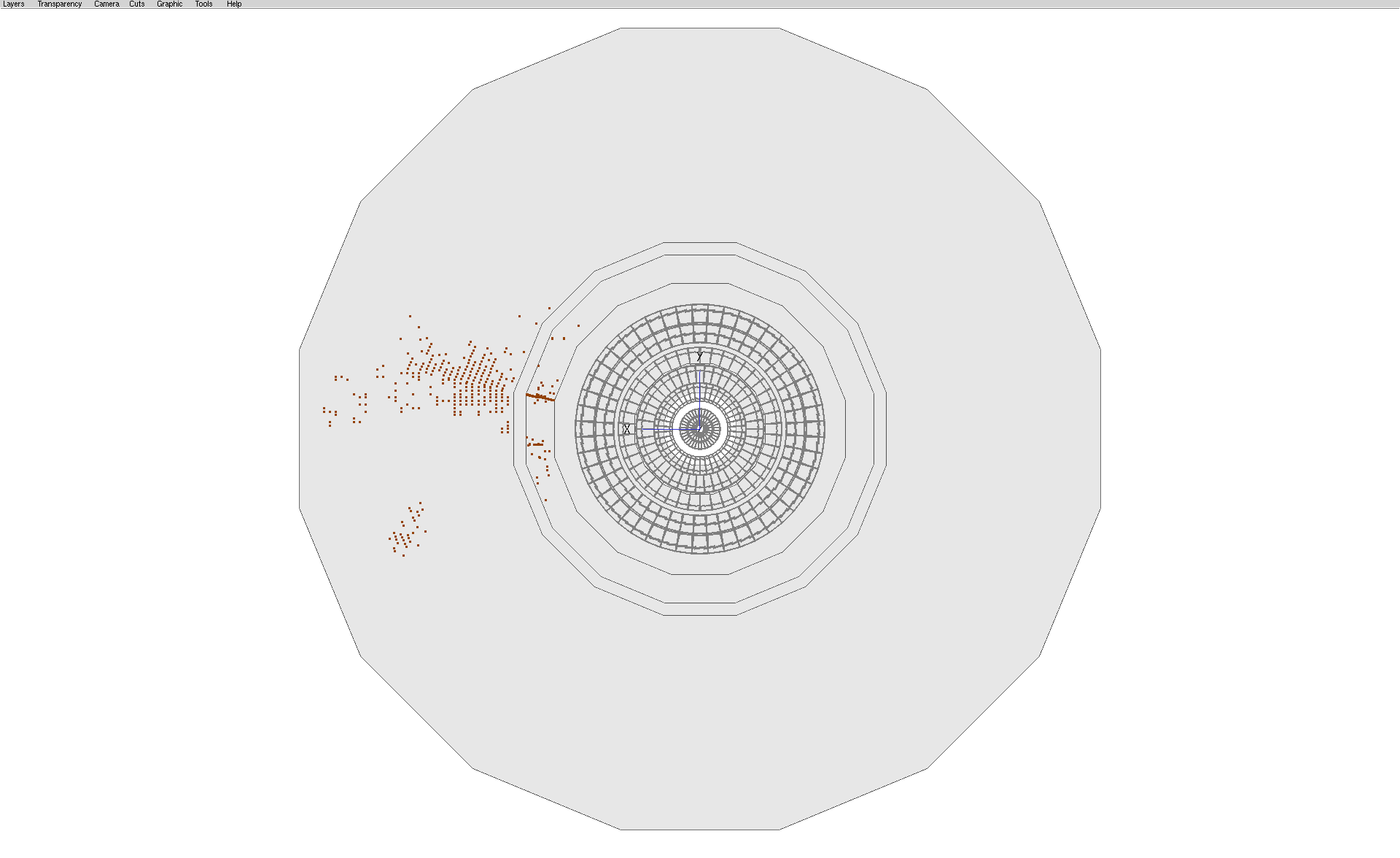}
        \caption{An event visualisation of a 10 GeV proton that enters the detector at $\theta=90^\circ$ and  $\phi=11.25^\circ$. The proton interacts early in the detector, and leaves distinctly separate clusters.}
        \label{fig:proton:phi11}
\end{figure}
\begin{figure}[htbp]
        \centering
        \includegraphics[width=\textwidth,trim={400pt 20pt 150pt 50pt},clip]{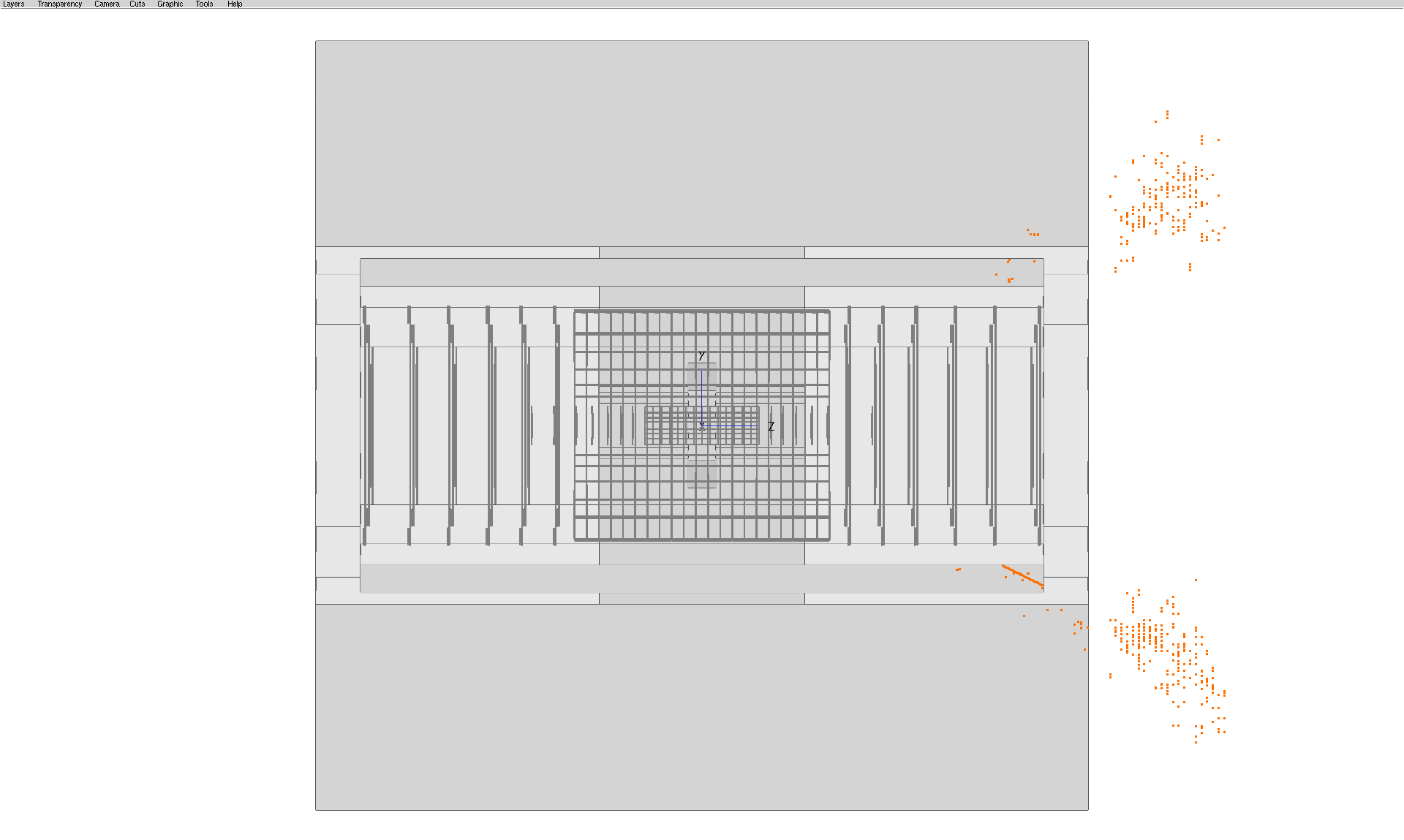}
        \caption{An event visualisation of a 10 GeV proton that enters the detector at $\theta=25^\circ$ and  $\phi=0^\circ$, which corresponds to the transition in between the barrel and hadronic endcap (the endcap is not displayed). The showers seems to leave deposition in two regions of the detector (bottom and top), but that is an artifact of how $\phi$ is displayed, as part of shower has low values, while the other part has values close to $360^\circ$.}
        \label{fig:proton:theta25}
\end{figure}
\clearpage

\section{Technical Validation}

Validation of all the files has been done using \texttt{validation/sanity\_checks.py} to inspect missing values, wrong ranges, and the number of entries.

\section{Code Availability}

The Open Data Detector can be found on GitLab in \cite{odd}. The XML configuration used in this simulation has been included in the step2point GitLab repository v.1.0.1~\cite{step2point}. The community standard key4hep software stack was used, with version 2025-05-29.

\subsubsection*{Simulation}
Based on one of the files from the continuous dataset for photons:

\begin{minted}[fontsize=\scriptsize,breaklines]{bash}
source /cvmfs/sw.hsf.org/key4hep/setup.sh -r 2025-05-29
source <ODD_installation_dir>/OpenDataDetector/install/bin/this_odd.sh
ddsim --steeringFile simulation/steer.py --compactFile simulation/ODD_xml/OpenDataDetector.xml --enableGun  --gun.distribution uniform  --gun.energy 10*GeV --gun.particle pi- --gun.phiMin 11.25*deg --gun.phiMax 11.25*deg --gun.thetaMin 90*deg --gun.thetaMax 90*deg --numberOfEvents 10 --outputFile ODD_piM_10ev_theta90deg_phi11.25deg_posX1250mmY248mmZ0mm_10GeV_edm4hep.root --gun.position "1250*mm 248*mm 0*mm"
ddsim --compactFile step2point/simulation/ODD_xml/OpenDataDetector.xml --steeringFile step2point/simulation/steer.py  --enableGun --gun.distribution uniform --gun.thetaMin 6*deg --gun.thetaMax 174*deg --gun.phiMin 0*deg --gun.phiMax 360*deg  --gun.momentumMin 0.1*GeV --gun.momentumMax 100*GeV --gun.particle gamma --numberOfEvents 10000 --random.seed 13361981 --gun.position "0 0 0" --outputFile step2point_ODD_gamma_0.1to100GeV_theta6to174deg_phi0to360deg_posX0Y0Z0_10000ev_file1_edm4hep.root
\end{minted}

The list of the random seeds used for the continuous dataset is given in Tab.~\ref{tab:seeds}. For the discrete part of the dataset the random seeds were attached to the metadata of the HDF5 files.

\begin{table}[htbp]
\centering
\begin{tabular}{c|c|c}
\hline
\textbf{particle} & \textbf{file ID} & \textbf{random seed for simulation} \\
\hline
gamma  & file1 & 13361981 \\
gamma  & file2 & 13361984 \\
gamma  & file3 & 13361987 \\
gamma  & file4 & 13361990 \\
gamma  & file5 & 13361993 \\
piM    & file1 & 13361982 \\
piM    & file2 & 13361985 \\
piM    & file3 & 13361988 \\
piM    & file4 & 13361991 \\
piM    & file5 & 13361994 \\
proton & file1 & 13360141 \\
proton & file2 & 13361983 \\
proton & file3 & 13361986 \\
proton & file4 & 13361989 \\
proton & file5 & 13361992 \\
\hline
\end{tabular}
\caption{Random seeds passed to the simulation for the continuous part of the dataset. This can be used to reproduce the files. The name of the file is \texttt{step2point\_ODD\_} \texttt{<PARTICLE>\_0.1to100GeV\_theta6to174deg\_phi0to360deg\_posX0Y0Z0\_10000ev\_<FILE\_ID>.h5}, where \texttt{<PARTICLE>} and \texttt{<FILE\_ID>} should be taken from the table.}\label{tab:seeds}
\end{table}

\subsubsection*{Translation}
Translation of simulation detailed information.

\begin{minted}[fontsize=\scriptsize
,breaklines]{bash}
python step2point/dataset/root2h5.py --input step2point_ODD_gamma_0.1to100GeV_theta6to174deg_phi0to360deg_posX0Y0Z0_10000ev_file1_edm4hep.root step2point_ODD_gamma_0.1to100GeV_theta6to174deg_phi0to360deg_posX0Y0Z0_10000ev_file1.h5
\end{minted}

\subsubsection*{Validation}
A validation of the files was performed to check the ranges and the lengths of values that are stored, as well as to check for all-zero or NaN values.

\begin{minted}[fontsize=\scriptsize
,breaklines]{bash}
python step2point/validation/sanity_check.py step2point_ODD_gamma_0.1to100GeV_theta6to174deg_phi0to360deg_posX0Y0Z0_10000ev_file1.h5
\end{minted}

\subsubsection*{\texttt{Cell\_id} encoding}
To help encoding the bitfield (\texttt{cell\_id}) from simulation steps, \texttt{utils/bitfield.py} can be used. For instance:

\begin{minted}[fontsize=\small]{python}
import sys,os
sys.path.append(os.path.abspath(os.path.join('.', 'utils')))
from bitfield import BitfieldCodec
codec = BitfieldCodec() # Default encoding string is CLD
import h5py
f = h5py.File("test/CLD_gamma_10GeV_posY2150mm_dirY1_10ev_sim_detailed_tchandler.h5","r")
codec.get(int(f['steps']['cell_id'][0])) # returns "layer" by default
codec.get(int(f['steps']['cell_id'][0]), "system")
\end{minted}

\subsubsection*{Visualisation}
This can be run on the original files (not the translation).

\begin{minted}[fontsize=\scriptsize,breaklines]{bash}
source /cvmfs/sw.hsf.org/key4hep/setup.sh -r 2025-05-29
source <ODD_installation_dir>/OpenDataDetector/install/bin/this_odd.sh
glced &
k4run step2point/simulation/event_display.py --inputFiles /eos/geant4/fastSim/ODD/step2point/step2point_ODD_gamma_0.1to100GeV_theta6to174deg_phi0to360deg_posX0Y0Z0_10000ev_file1_edm4hep.root --compactFile step2point/simulation/ODD_xml/OpenDataDetector.xml
\end{minted}

\begin{figure}[htbp]
        \centering
        \includegraphics[width=1\textwidth,trim={30pt 10pt 100pt 50pt},clip]{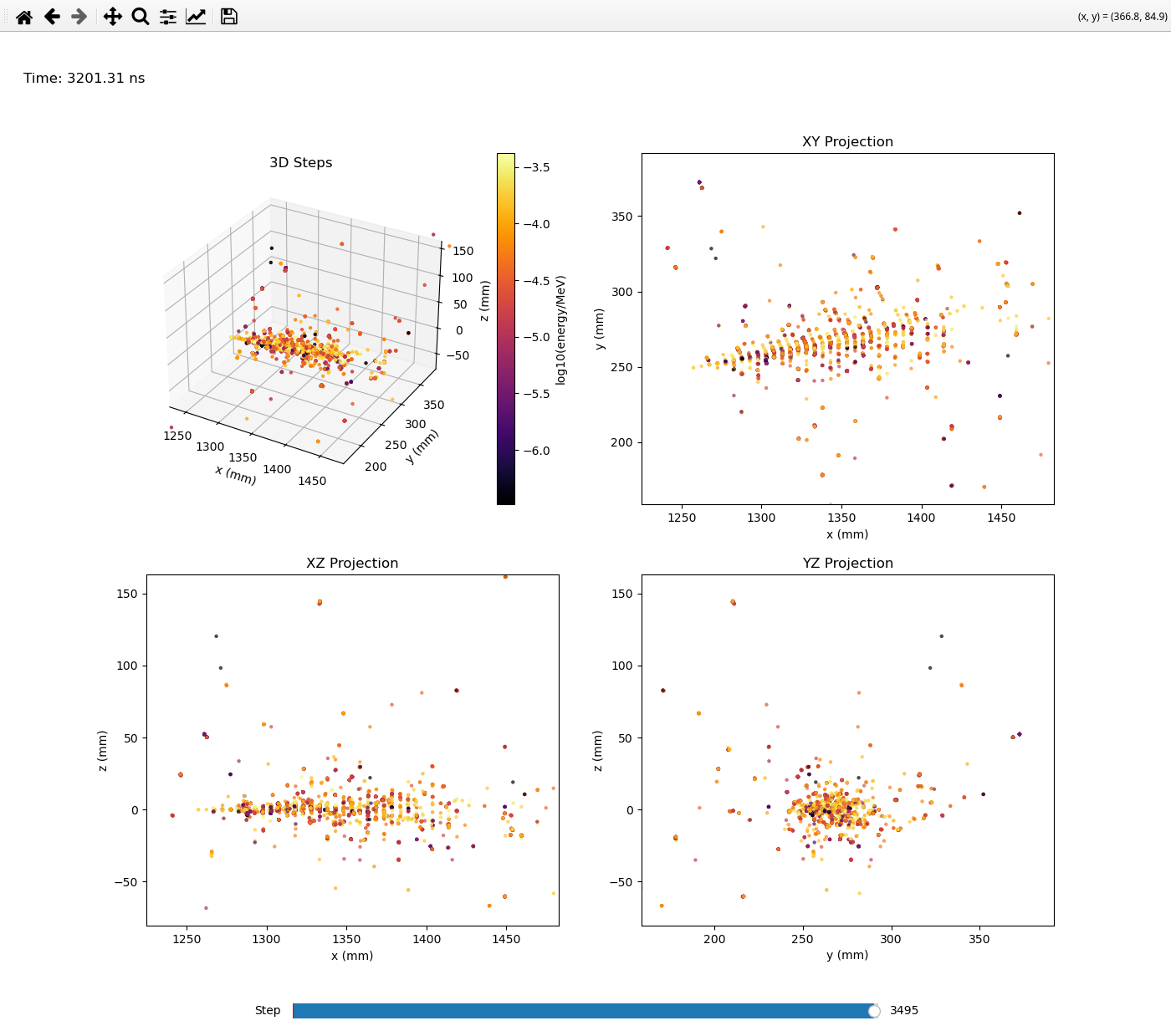}
        \caption{A simple visualisation of the 10 GeV photon (the last frame of an animated gif). The XY projection shows clearly how the energy is deposited on layers in the corner of the modules of the detector.}
        \label{fig:animation}
\end{figure}

\noindent On the translated files we can run a simplified visualisation, without the detector definition, as can be seen in Fig.~\ref{fig:animation}.

\begin{minted}[fontsize=\scriptsize,breaklines]{bash}
python step2point/visualisation/animate_event.py step2point_ODD_gamma_10GeV_theta90deg_phi0deg_posX1250Y0Z0_1000ev.h5 1 --color_by=energy --auto_interval
\end{minted}

\section{Data Availability}

The dataset is available on Zenodo~\cite{step2point_zenodo}.

\bibliographystyle{unsrt_custom}
\bibliography{references}

\end{document}